# Amplitude-based echo-mode speed-of-sound tomography

Michael Jaeger[1,*], Marina Eckermann[1]

[1]Institute of Applied Physics, University of Bern, Switzerland

*Corresponding author: michael.jaeger@unibe.ch

## Abstract

The tissue's speed of sound (SoS) is a promising marker for non-invasive and zero-dose diagnosis of diseases such as cancer or liver steatosis. Computed ultrasound tomography in echo mode (CUTE) retrieves spatially resolved SoS using pulse-echo ultrasound, by analysing the variation of echoes when probed with varying pairs of transmit- and receive-steering angles. This analysis conventionally focuses on the echo phase shift, which is interpreted as differential time-of-flight (dToF) and related to SoS. While promising results have been obtained, the trade-off between beamwidth and angular aperture not only limits spatial resolution but also leads to artefactual noise in media with short-scale SoS variations. To overcome this limitation, we propose an alternative approach: for each echo, instead of reducing the angle-dependent variations to a single parameter (the dToF), we retrieve aberration profiles with sub-beamwidth resolution by inverting a model that relates these profiles to the echo amplitudes detected for all angle pairs. We demonstrate in simulations that this amplitude-based (a-CUTE) approach results in substantially improved spatial resolution and reduced artifacts compared to dToF-CUTE. Physical phantom and preliminary *in vivo* results confirm the feasibility of this novel method in physical data.

## 1. Introduction

Diagnostic ultrasound (US) is a cornerstone of modern medicine due to its wide availability, portability, and non-ionizing radiation. On the downside, its standard brightness (B)-mode display of echo intensity remains largely qualitative and operator-dependent, often limiting diagnostic accuracy and reproducibility. The tissue's speed of sound (SoS) offers a promising quantitative biomarker for disease-related changes to tissue composition. For this reason, the scientific community strives to develop techniques to derive the SoS from US measurements. One such approach is through-transmission ultrasound tomography (UT): conceptually analogous to X-ray tomography, the analysis of US wavefronts transmitted through the tissue along a variety of propagation angles (ideally spanning 180°) provides projections of inverse SoS (i.e., slowness), which allow us to retrieve spatially resolved maps of SoS. Historically, the time of flight (ToF) of signals was interpreted as line integrals of slowness along ray paths [1–8], however, the finite width of US beams due to diffraction limits spatial resolution, and self-interference after scattering at sub-wavelength structures introduces ambiguity in ToF identification. These challenges have been largely solved by retrieving the slowness from the detected signal amplitude (magnitude and phase) as opposed to the ToF, allowing to account for diffraction and interference [9–11]. Excellent spatial resolution (on the scale of the wavelength) and contrast have thus been demonstrated in the breast [9,12–14] and recently in the adult leg and full-body imaging of piglets [15,16]. However, UT is limited to transparent tissues that are accessible from all sides, and it requires heavy stand-alone equipment (circular or rotating detection geometry with water bath for acoustic coupling) lacking the portability, flexibility, and wide availability of echo US.

With the goal to combine this novel quantitative modality with the portability, flexibility, and wide availability of echo US, the feasibility of spatial mapping of SoS using handheld echo US has been demonstrated [17–30] asides methods to estimate average SoS [31–34]. Among them, computed ultrasound tomography in echo mode (CUTE) [17,18,20,22,27] relies on probing echoes with varying combinations of steered transmitted (Tx) and receiving (Rx) ultrasound beams. The differential ToF (dToF) along varying steering angles—measured as an echo phase shift—is related to line integrals of slowness along the beam paths from which the SoS is inverted. While promising results were obtained in the context of diagnosing fatty liver disease [35] and breast cancer [36,37], echo mode SoS imaging shares the limitation of travel time UT: the variation of echo amplitude and phase is compressed to a single metric, the dToF, so that information on short-scale aberrations of the interrogating fields below the beamforming resolution is lost, limiting spatial and contrast resolution [38].

Inspired by UT, we hypothesize that a direct analysis of the radio-frequency echo amplitude as opposed to echo shift circumvents the current limitation of CUTE. Towards this goal, this exploratory study investigates the feasibility of amplitude-based (a-)CUTE. We relate the beamformed echo amplitudes detected with varying Tx/Rx angle pairs to the transversal wavefront profiles and transversal profiles of reflectivity. We show that this relation can be inverted given sufficient angle pairs, allowing to retrieve wavefront profiles and thus SoS maps with strongly improved lateral resolution and contrast compared to classical dToF-CUTE.

## 2. Theory

To help the lecture of this section, Table I at the end of the manuscript provides a glossary summarizing the mathematical symbols.

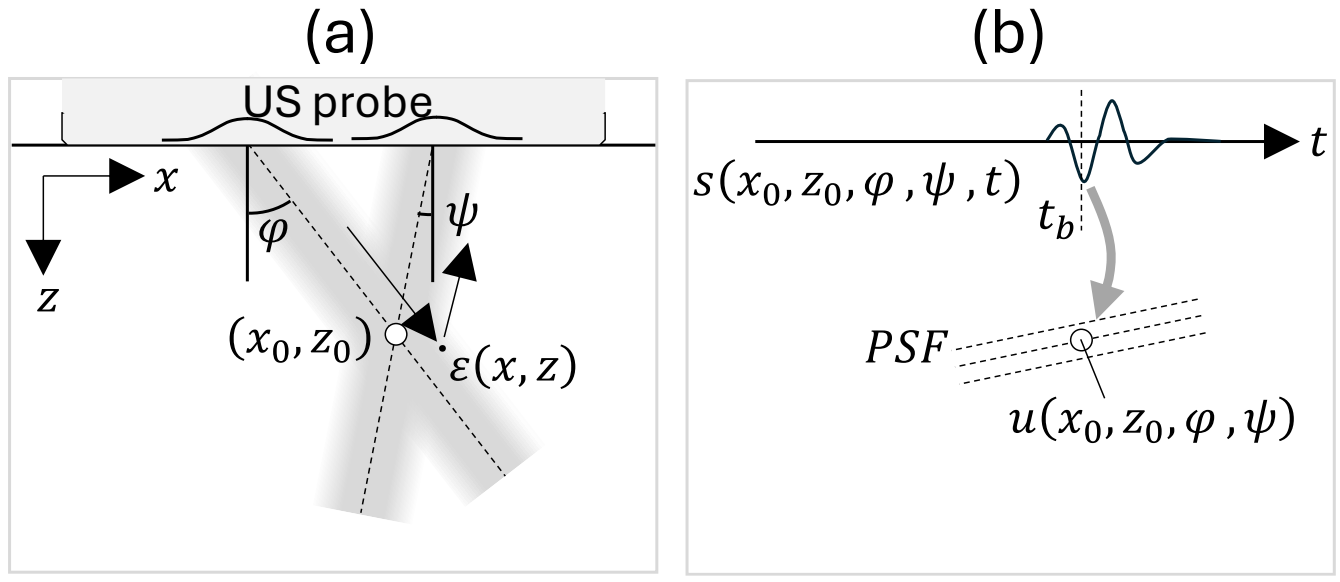

**Fig. 1:** (a) Sketch of the acquisition geometry. A linear array probe is placed on top of a 2D tissue volume. It probes the reflectivity around a pivot $(x_0, z_0)$ by transmitting an US pulse along a collimated beam (angle $\varphi$) and receiving US along a second collimated beam (angle $\psi$). A point with reflectivity $\varepsilon(x,z)$ leads to an echo. (b) Echoes are seen as oscillations in the signal $s(x_0, z_0, \varphi, \psi, t)$. The amplitude at the anticipated round-trip time $t_b$ is assigned to the pivot location in the beamformed image $u(\ldots)$. This value corresponds to an integral over $\varepsilon(x,z)$, weighted by the point-spread function (PSF).

### 2.1 Physical model of beamformed ultrasound echoes

We consider the scenario depicted in **Fig. 1a**, where a linear array probe is placed on top of a 2D volume with coordinates $x$ (parallel to the array axis/tissue surface, transversal direction) and $z$ (perpendicular to the array axis, axial direction). In CUTE, we beamform complex radio frequency (CRF)-mode images using varying transmit (Tx)/receive (Rx) steering angle pairs: for each angle pair, we probe individual nodes with coordinates $(x_0, z_0)$ on a pre-defined Cartesian grid by transmitting a beam with angle $\varphi$ (relative to $z$) and receive along a beam with angle $\psi$ so that the two beams intersect at $(x_0, z_0)$. We henceforth refer to $(x_0, z_0)$ as "pivot". Even though the details of implementation can differ, for didactic simplicity one may assume that the Tx beam is generated physically via Tx delays and apodization weights of individual elements, and the Rx beam is implemented via Rx delays and apodization weights in delay-and-sum of signals detected on individual elements. We assume that the apodization weights and delays are chosen to create collimated Gaussian beams. **Fig. 1a** sketches a pair of intersecting Tx/Rx beams.

Conceptually following the 1st order Born approximation, the contribution to the detected signal $s(x_0, z_0, \varphi, \psi, t)$ of a point reflector with infinitesimal size $dx, dz$ located at $(x, z)$ can be expressed as:

$$s(x_0, z_0, \varphi, \psi, t) = dx\, dz\, \varepsilon(x,z) \cdot \int dt'\; p_\varphi^{(x_0,z_0)}(x,z,t') p_\psi^{(x_0,z_0)}(x,z,t-t'), \quad (1)$$

where $p_{\varphi/\psi}^{(x_0,z_0)}(x,z,t)$ are the spatio-temporal wavefields of the two beams, and we assume that the physical property responsible for echo generation is a quantity $\varepsilon(x,z)$ henceforth called "reflectivity". The wavefields are assumed to be complex-valued with a one-sided temporal frequency spectrum. For a spatially distributed $\varepsilon(x,z)$, the right-hand side of **Eq.1** is integrated over $(x,z)$. The CRF-mode image amplitude $u(x_0, z_0, \varphi, \psi)$ at the location of the pivot is defined as the value of $s(x_0, z_0, \varphi, \psi, t = t_b)$, where $t_b(x_0, z_0, \varphi, \psi)$ is the anticipated echo round-trip time for an *a priori* beamforming SoS $c_b$:

$$u(x_0, z_0, \varphi, \psi) = s\big(x_0, z_0, \varphi, \phi, t_b(x_0, z_0, \varphi, \psi)\big) = \iint dx\, dz\, \varepsilon(x,z)\, h_{\varphi,\psi}^{(x_0,z_0)}(x,z)\ , \quad (2)$$

where $h_{\varphi,\psi}^{(x_0,z_0)}(x,z) = \int dt'\; p_\varphi^{(x_0,z_0)}(x,z,t') \cdot p_\psi^{(x_0,z_0)}(x,z,t_b - t')$

is the point-spread function (PSF). The procedure is illustrated for a single pivot in **Fig. 1b**. It is repeated for all pivots on a pre-defined 2D grid covering the desired field-of-view (FoV).

The wavefields $p_{\varphi/\psi}^{(x_0,z_0)}(x,z,t)$ are determined by the parameters $(x_0, z_0, \varphi, \psi)$ (more specifically, by the per-element Tx/Rx delays and apodization weights), as well as by the interaction with the spatial distribution of the true SoS $c(x,z)$. We assume that aberration-related propagation delays are smaller than the temporal pulse length such that the disturbed wavefields can be written in terms of the undisturbed fields $P_{\varphi/\psi}^{(x_0,z_0)}(x,z,t)$ (for $c = c_b$) and a phase rotation given by aberration profiles $\theta_{\varphi/\psi}(x,z)$:

$$p_{\varphi/\psi}^{(x_0,z_0)}(x,z,t) = e^{i\theta_{\varphi/\psi}(x,z)} P_{\varphi/\psi}^{(x_0,z_0)}(x,z,t)\ . \quad (3)$$

Thereby, the aberration profiles can take complex values, so that—in addition to a phase delay (real part)—they can describe the amplitude modulation due to self-interference (imaginary part), even though the latter will not be used in this study. Redefining the spatial coordinates relative to $(x_0, z_0)$, $(x', z') = (x - x_0, z - z_0)$, **Eq. 2** becomes:

$$u(x_0, z_0, \varphi, \psi) = \iint dx'\, dz'\, \varepsilon(x', z') e^{i\theta_\varphi(x', z')} e^{i\theta_\psi(x', z')}\, H_{\varphi,\psi}^{(x_0,z_0)}(x', z'), \quad (4)$$

with $H_{\varphi,\psi}^{(x_0,z_0)}(x', z') = \int dt'\; P_\varphi^{(x_0,z_0)}(x', z', t') \cdot P_\psi^{(x_0,z_0)}(x', z', t_b - t')$ .

The undisturbed PSF, $H_{\varphi,\psi}^{(x_0,z_0)}(x', z')$, is illustrated in **Fig. 2a**. It is a wavelet with a spatial oscillation along the mid-angle between $\varphi$ and $\psi$, with oscillation period and axial profile determined by the centre frequency and length of the impulse response, respectively, and a transversal profile determined by the product of the transversal beam profiles.

## 2.2 Aberration model for a-CUTE

In standard dToF-CUTE we assumed that the aberration profiles are constant across the width of the undisturbed PSF and can thus be factored out of the integral in **Eq. 4**. In this study, we go beyond standard CUTE and account for the fact that short-scale variations of SoS below the transversal beamwidth lead to transversal variations of phase (and amplitude) across the beam profile. In the following we demonstrate that—around a given pivot $(x_0, z_0)$—it is possible to retrieve the spatially varying aberration profiles from the collection of all CRF-mode image amplitudes sampled with varying Tx/Rx pairs. For this purpose, we make a model for the undisturbed PSF where we assume that the angles $(\varphi, \psi)$ are small so that the PSF is roughly aligned with the $x$-axis:

$$H_{\varphi,\psi}^{(x_0,z_0)}(x', z') = W_\varphi(x') W_\psi(x') \cdot T^2(z') \cdot e^{i(k_{x,\varphi} + k_{x,\psi})x'} \cdot e^{2ikz'}, \quad (5)$$

where $W_\varphi(x') W_\psi(x')$ is the product of the transversal beam profiles, $T(z')$ is the axial envelope determined by the temporal envelope of the impulse response, $k = 2\pi\, f_c / c_b$ is the spatial frequency corresponding to the centre frequency $f_c$ and $k_{x,\varphi/\psi} = k \sin \varphi/\psi$ describe the transversal oscillations of the two beams. The small-angle approximation is contained in (i) setting $k_z = k$, and (ii) the independence of $T(z')$ from $x'$. The latter is reasonable for angles where the tails of the PSF rotate relative to the $x$-axis by less than the axial width $\delta z$ of the profile $T^2(.)$ (illustratedin **Fig. 2a**). If we furthermore assume that the aberration profiles do not vary along $z'$ within the axial width of $T^2(.)$, **Eq. 4** combined with the PSF model **Eq. 5** becomes:

$$u(x_0, z_0, \varphi, \psi) = \int dx'\; W^2(x') \cdot e^{i(k_{x,\varphi} + k_{x,\psi})x'} \cdot e^{i(\theta_\varphi(x') + \theta_\psi(x'))} \cdot \tilde{\varepsilon}(x'), \quad (6)$$

with $\tilde{\varepsilon}(x') = \int dz'\; T^2(z') \cdot \varepsilon(x', z') \cdot e^{2ikz'}$ .

The small-angle approximation thus allows us to replace the integral of $\varepsilon$ over $z'$ by an effective reflectivity $\tilde{\varepsilon}(x')$ that is independent of $(\varphi, \psi)$.

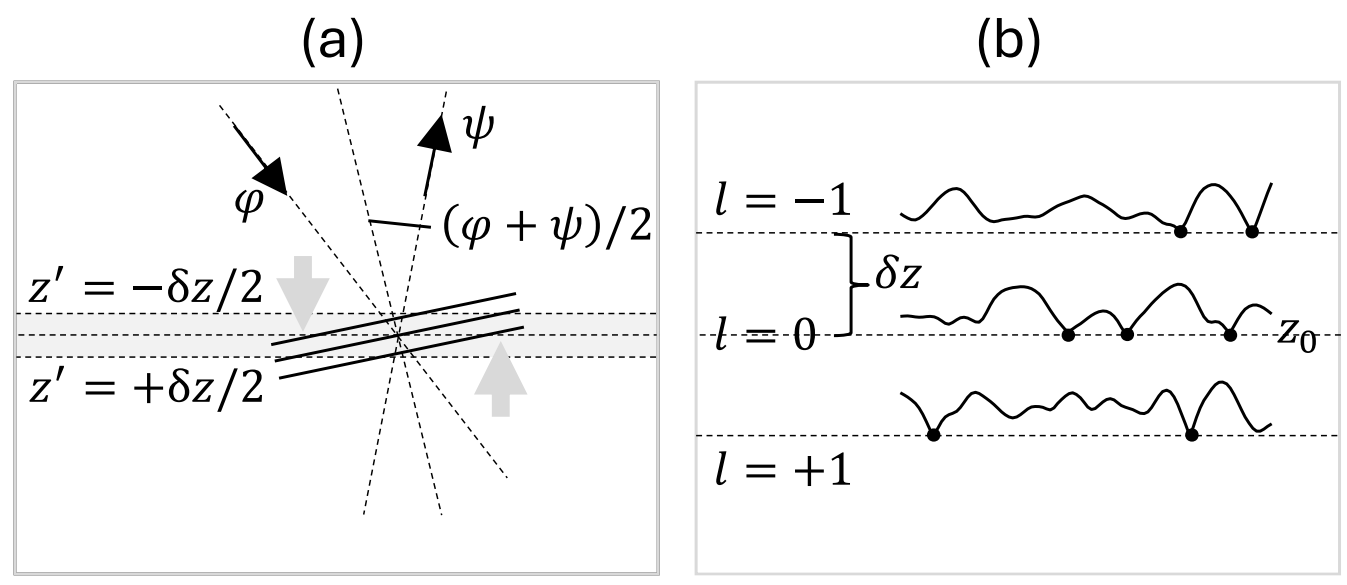

**Fig. 2:** (a) Sketch of the diameter $\pm\delta z/2$ of the approximate depth layer over which the PSF (defined in Eq. 5) integrates reflectivity in Eq. 6. Even though the PSF (solid black lines indicate isochrones) is aligned with this layer only if the mid-angle $(\varphi + \psi)/2$ is 5parallel to the $z$-axis, we assume that rotation of the lateral tails of the PSF (gray arrows) out of this layer is negligible. (b) Illustration of multiple layers of reflectivity (absolute value shown) spaced by $\delta z$ around $z_0$. If amplitudes are statistically independent between layers, near-zero values (dots) are found at different positions in the different layers.

For an individual $(x_0, z_0)$, **Eq. 6** links the measurements $u(x_0, z_0, \varphi, \psi)$ to the unknown effective reflectivity $\tilde{\varepsilon}(x')$ and aberration profiles $\theta_{\varphi/\psi}(x')$. In this study, we demonstrate that—provided measurements are performed for a sufficient number of angle pairs—the unknowns can be retrieved. For this purpose, we discretize **Eq. 6**: the $x'$ coordinate is sampled by $N$ discrete values spaced by $\delta x$, $\{x_n\;,\; n \,\epsilon\, [1{:}\,N]\}$, covering the interval where $W_n^{\,2} = W^2(x_n)$ deviates substantially from zero (the "support"), and angles $(\varphi, \psi)$ are drawn from a set of $M$ discrete values spaced by $\delta\phi$, $\{\phi_m\;,\; m \,\epsilon\, [1{:}\,M]\}$. The discretized **Eq. 6** reads:

$$u_{m1,m2} = \sum_{n=1}^{n=N} W_n^{\,2} \cdot e^{i(k_{m1} + k_{m2})x_n} \cdot \Theta_{m1,n} \cdot \Theta_{m2,n} \cdot \tilde{\varepsilon}_n, \quad (7)$$

where we have simplified the notation of the phase rotation by defining $\Theta_{m,n} \stackrel{\text{def}}{=} e^{i\theta_{\phi m}(x_n)}$. Given Tx/Rx reciprocity, $\frac{1}{2}M(M+1)$ out of the $M^2$ measurements $u_{m1,m2}$ are independent. Our study departs from the hypothesis that the $\tilde{\varepsilon}_n$ and $\Theta_{m,n}$ can in principle be retrieved if the number of independent measurements is larger or equal to the number $(M+1)N$ of unknown values ($\tilde{\varepsilon}_n$, $\Theta_{m,n}$), which is the case if $M \geq 2N$. In the following we propose an iterative solution for the retrieval of the unknown values ($\tilde{\varepsilon}_n$, $\Theta_{m,n}$) from the measurements $u_{m1,m2}$.

2.3 Retrieval of reflectivity for given aberration profiles

**Eq. 7** is linear in $\tilde{\varepsilon}_n$, thus, assuming the $\Theta_{m,n}$ are known, it can be written as a matrix-vector product:

$$\mathbf{u} = \mathbf{\Theta}\boldsymbol{\varepsilon}, \tag{8}$$

where $\mathbf{u}$ is a vector with length $M^2$ containing the measurements $u_{m1,m2}$, $\boldsymbol{\varepsilon}$ is a vector with length $N$ containing the values $\tilde{\varepsilon}_n$, and $\mathbf{\Theta}$ is a matrix with dimensions $M^2$ by $N$ containing the values $W_n{}^2 \cdot e^{i(k_{m1}+k_{m2})x_n} \cdot \Theta_{m1,n} \cdot \Theta_{m2,n}$. The unknowns $\boldsymbol{\varepsilon}$ can be retrieved from the measurements $\mathbf{u}$ via a pseudo inverse $\mathbf{\Theta}^\dagger$ of $\mathbf{\Theta}$:

$$\boldsymbol{\varepsilon} = \mathbf{\Theta}^\dagger \mathbf{u}, \text{ with } \mathbf{\Theta}^\dagger \stackrel{\text{def}}{=} (\mathbf{\Theta}^*\mathbf{\Theta} + \gamma_\varepsilon^2 \mathbf{I})^{-1}\mathbf{\Theta}^*, \tag{9}$$

where the asterisk denotes conjugate transpose, $\mathbf{I}$ denotes the identity matrix, and $\gamma_\varepsilon^2$ is a regularisation parameter. Note that one could in principle limit the elements in $\mathbf{u}$ and thus the number of rows in $\mathbf{\Theta}$ to the $\frac{1}{2}M(M+1)$ independent measurements. For physical data, however, including redundant measurements provides an advantageous averaging effect.

2.4 Update of phase delays for given reflectivity

Unfortunately, **Eq. 7** is not linear in $\Theta_{m,n}$ (it contains a product of two $\Theta_{m,n}$ for $m = m1$ and $m2$). However, assuming the $\tilde{\varepsilon}_n$ are known, an updated guess of $\Theta_{m,n}$ can be obtained from the previous guess. For this we write the forward problem in **Eq. 7** as $M$ separate problems: for each $m2$, the measurements $u_{m1,m2}$ (function of $m1$) are related to a vector $\boldsymbol{\theta}_{m2}$ with length $N$ containing the $\Theta_{m2,n}$ for all $n$. In matrix notation:

$$\mathbf{u}_{m2} = \mathbf{E}_{m2}\boldsymbol{\theta}_{m2}, \tag{10}$$

where $\mathbf{u}_{m2}$ is a vector with length $M$ containing the measurements $u_{m1,m2}$ for fixed $m2$, and $\mathbf{E}_{m2}$ is a matrix with dimensions $M$ by $N$ that contains the values $W_n{}^2 \cdot \tilde{\varepsilon}_n \cdot e^{i(k_{m1}+k_{m2})x_n} \cdot \Theta_{m,n}$ computed from the previous guess of $\theta_{m,n}$. Analogue to **Eq. 9**, the updated $\boldsymbol{\theta}_{m2}$ can be retrieved from the measurements $\mathbf{u}_{m2}$ via a pseudo inverse $\mathbf{E}_{m2}^\dagger$ of $\mathbf{E}_{m2}$ with a regularisation parameter $\gamma_\theta^2$:

$$\boldsymbol{\theta}_{m2} = \mathbf{E}_{m2}^\dagger \mathbf{u}_{m2}, \text{ with } \mathbf{E}_{m2}^\dagger \stackrel{\text{def}}{=} (\mathbf{E}_{m2}^*\mathbf{E}_{m2} + \gamma_\theta^2 \mathbf{I})^{-1}\mathbf{E}_{m2}^*. \tag{11}$$

By repeating the retrieval of $\boldsymbol{\theta}_{m2}$ for each $m2$, all updated $\Theta_{m,n}$ are obtained. Note that the relative contribution of the values of $\Theta$ and $\tilde{\varepsilon}$ to $u$ is ambiguous. To avoid this ambiguity, after completing the update, we normalize (divide) $\Theta_{m,n}$ for each $n$ by its root-mean-square (RMS) along $m$. This convention does not influence the value of the aberration delay (exponent of $\Theta_{m,n}$) but preserves angle-dependent variations of the absolute value of $\Theta_{m,n}$ that are important in the context of attenuation imaging.

2.5 Alternating retrieval of reflectivity and phase delays

So far, we assumed that either the phase delays or the reflectivity values are known. To retrieve both, we alternatingly employ **Eqs. 9** and **11**, starting from an initial guess of $\Theta_{m,n}$. The procedure for retrieving reflectivity and aberration delays around a chosen pivot $(x_0, z_0)$ is thus as follows:

(a) Beamform the measurements $u_{m1,m2}$ for all Tx/Rx angle pairs $(\varphi, \psi) = (\phi_{m1}, \phi_{m2})$.
(b) Initialize the values of $\Theta_{m,n}$ to 1.
(c) Retrieve $\tilde{\varepsilon}_n$ from $u_{m1,m2}$ and $\theta_{m,n}$ via $\boldsymbol{\varepsilon} = \mathbf{\Theta}^\dagger \mathbf{u}$.
(d) For each $m2$, retrieve the updated $\Theta_{m2,n}$ from $u_{m1,m2}$, the previous $\Theta_{m,n}$, and $\tilde{\varepsilon}_n$ by $\boldsymbol{\theta}_{m2} = \mathbf{E}_{m2}^\dagger \mathbf{u}_{m2}$.
(e) Divide $\Theta_{m,n}$ for each $n$ by its RMS along $m$.
(f) Iterate through (c) to (e) a number of times.
(g) Derive the phase delays $\theta_{\phi_m}(x_n) = \arg \Theta_{m,n}$.

2.6 Refined retrieval of phase delays

We will show in the Results section that this basic retrieval approach leads to noisy $\Theta_{m,n}$. The reason is that the reflectivity $\tilde{\varepsilon}_n$ has values near zero (illustrated in **Fig 2b**) in points where diffuse echoes interfere destructively, so that $\Theta_{m,n}$ cannot be reliably retrieved in these points. If we assume that the phase delays $\theta_{\varphi/\psi}(x', z)$ vary along $z$ more slowly than the spatial scale $\delta z$ of axial variations of $\tilde{\varepsilon}$, we can retrieve $\Theta_{m,n}$ simultaneously from multiple measurements performed at $L$ depths spaced by $\delta z$, i.e., from $u(x_0, z_0 + l \cdot \delta z, \varphi, \psi), l \epsilon [0.5 : L - 0.5] - L/2$. This must go along with retrieving $L$ layers of reflectivity, i.e., $\tilde{\varepsilon}(x', l \cdot \delta z, \varphi, \psi)$. If the reflectivity profiles at different depths are statistically independent, the probability of obtaining near-zero values at the same $x'$ for all depths is small (illustrated in **Fig. 2b**), so that estimation errors of $\theta_{\varphi/\psi}(x')$ are reduced. For this purpose, **Eq. 6** and **Eq. 7** are re-formulated to:

$$u(x_0, z_0 + l \cdot \delta z, \varphi, \psi) = \int dx' \; W^2(x') \cdot e^{i(k_{x,\varphi}+k_{x,\psi})x'} \cdot e^{i(\theta_\varphi(x')+\theta_\psi(x'))} \cdot \tilde{\varepsilon}(x', l \cdot \delta z) \tag{12}$$

and

$$u_{m1,m2,l} = \sum_{n=1}^{n=N} W_n{}^2 \cdot e^{i(k_{m1}+k_{m2})x_n} \cdot \Theta_{m1,n} \cdot \Theta_{m2,n} \cdot \tilde{\varepsilon}_{n,l} \; . \tag{13}$$

The forward model and inversion in **Eq. 8** can be formulated for the different $l$ separately. Because the matrix $\mathbf{\Theta}$ does not depend on $l$, the same pseudo-inverse can be applied in **Eq. 9** to retrieve vectors $\boldsymbol{\varepsilon}_l$ from vectors $\mathbf{u}_l$ which contain the values $\tilde{\varepsilon}_{n,l}$ and $u_{m1,m2,l}$ for one $l$ at a time. In **Eq. 10**, however, the measurements $\mathbf{u}_l$ for different $l$ are related to the same $\boldsymbol{\theta}_{m2}$. To simultaneously retrieve $\boldsymbol{\theta}_{m2}$ from all $\mathbf{u}_l$, the forward models for all $l$ must thus be combined in a single matrix $\mathbf{E}$ with dimensions $M^2L$ by $N$, and $\mathbf{u}_{m2}$ is a vector with length $ML$ combining the measurements $\mathbf{u}_l$.

2.7 Retrieval of aberration profiles, full field-of-view

The measurements $u_{m1,m2}$ probe transversal profiles (along $x'$) of reflectivity and aberrations around a specific pivot $(x_0, z_0)$. Thereby, the width of the profile $W^2(x')$ limits the $x'$-interval ("support") inside which the reflectivity and aberration profiles have a notable influence on the measurement. Consequently, the retrieval can be limited to nodes $x_n$ inside this support, and the number $N$ of nodes is determined by the ratio between the support length and the transversal grid spacing $\delta x$. Given a desired FoV, one could in principle choose the $W^2(x')$ so that the full width of the FoV can be retrieved in one go, however, this also results in a large $N$, and—in turn—in a large number of required angles $M \geq 2N$. This can also be seen from the perspective of a transversal Fourier transform: the required radius $\phi_{max}$ of the angular range $[-\phi_{max}: \delta\phi: \phi_{max}]$ (more precisely, the maximum transversal frequency $k_{x,\varphi}$) is inversely related to the desired transversal grid spacing $\delta x$, and thus fixed for a given spacing. The required angular spacing $\delta\phi$ (more precisely, the spacing $\delta k$ of transversal frequencies), however, is inversely related to the width of $W^2(x')$ (see Results section) which thus determines the number of required angles. Because reducing the beamwidth not only reduces $N$, but also $M$, it over-proportionally reduces the computational cost for constructing and inverting the matrices in **Eq. 9** and **11**. In addition, the smaller the width of $W^2(x')$, the better the small angle approximation is fulfilled. Like in standard dToF-CUTE, we thus minimize the width of $W^2(x')$ while maintaining collimated beams.

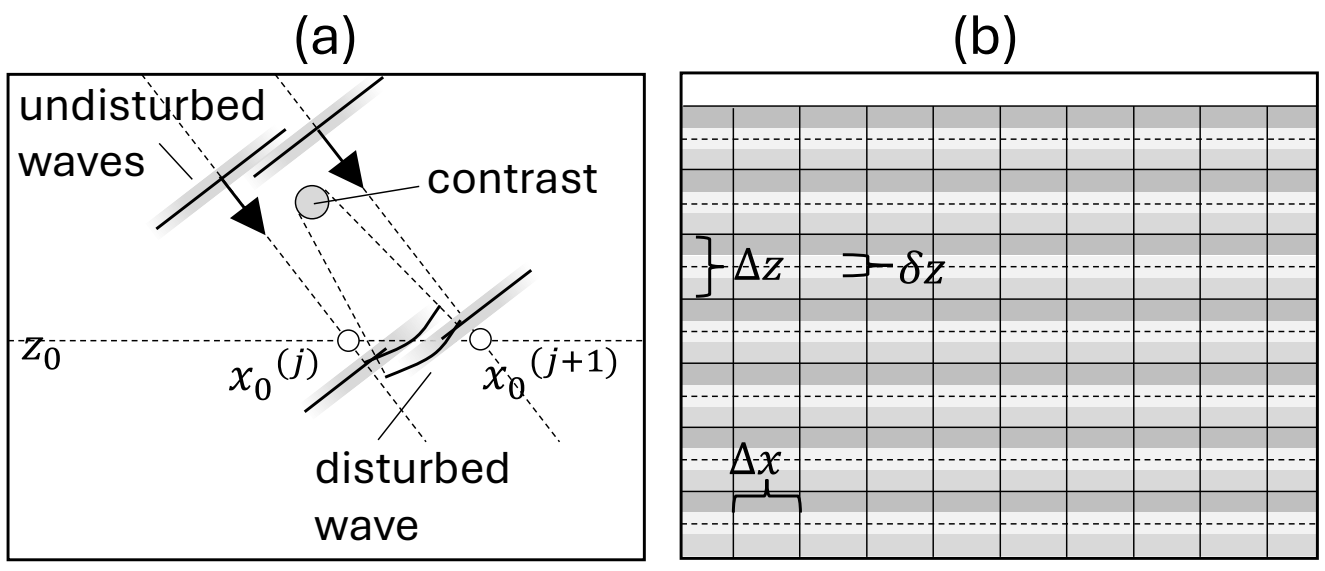

**Fig. 3:** (a) With collimated beams, the disturbed wavefronts for two beams pointing at pivots with different $x_0$ do not depend on the position of the beam axes relative to a SoS contrast. (b) Sketch of the beamforming grid, with transversal $\Delta x$ and axial spacing $\Delta z$ between pivots $(x_0, z_0)$, as well as the spacing $\delta z$ between different reflectivity layers.

To cover the full width of the FoV limited by the full probe aperture, the reflectivity and phase delay profiles must be retrieved from measurements performed at multiple $x_0$, spaced by a $\Delta x$ that allows a partial overlap of the support of adjacent beams. In this study we assume that the beams are truly collimated, i.e., the angle of incidence of the wavefield on aberrating structures does not depend on $x_0$ (for a focused beam, it would substantially change) and thus the transversal profile of aberrations (in absolute coordinates, $x$) does not depend on $x_0$, see **Fig. 3a**. This allows us to perform for each angle individually a weighted average of aberration profiles obtained for different $x_0$. Finally, to cover the full depth range of the desired FoV, the procedure is repeated for multiple $z_0$ spaced by $\Delta z$. Thereby, $\Delta z$ contains multiple $\delta z$, and the ratio between the two determines the number $L$ of independent layers probed for each $z_0$. The geometry of the full beamforming grid is illustrated in **Fig. 3b**.

## 3. Results and Discussion

3.1 Illustration of a-CUTE in simple simulations

In a first step, we illustrate the various concepts and steps of a-CUTE in a simplified simulation where the measurements are simulated using the forward model **Eq. 16**. That way, the results represent what is most optimistically achievable when the measurement process is accurately represented by the model used for the inversion. **Fig. 4a** illustrates the assumed scenario, where the sample consists of a row of five cylinders of $2\,mm$ diameter placed at a constant depth, with equidistant transversal spacing of $5\,mm$. The settings of the US acquisition mimic the actual US system used in the physical results: a linear array with an aperture length of $38\,mm$ (128 elements at $0.3\,mm$ pitch) and $5\,MHz$ centre frequency. With $c_b = 1.5\,\text{mm}/\mu\text{s}$, the wavelength $\lambda_c$ corresponding to the centre frequency is $0.3\,mm$. For didactic simplicity we limit our simulations to a single pivot depth $z_0$, where we discretize the ground truth $\tilde{\varepsilon}_l(x)$ and $\theta_\phi(x)$ on a $38\,mm$ aperture with a $\delta x = 0.075\,mm$ ($0.25\,\lambda_c$) resulting in 509 nodes. The grid resolution for the simulation was chosen smaller than for the retrieval (see below) to avoid the inverse crime. For each node independently, we draw random samples of the real and imaginary part of reflectivity from a normal distribution. For the ground truth $\theta_\phi(x)$ we assume that each cylinder leads to a Gaussian phase delay profile, with $1\,mm$ $1/e^2$ radius and $1\,rad$ peak value. The combined profile is transversally shifted relative to the cylinder positions depending on $\phi$, by $5\,mm \cdot \tan\phi$, to mimic a scenario where $z_0$ is located $5\,mm$ below the cylinders. **Fig. 4b** shows the ground truth phase delay as a function of $x$ for $\phi$ ranging within $\pm 40°$. Note that it is irrelevant whether the Gaussian profile is a realistic model for the phase delay as it only serves an illustrative purpose. Note also that—in a first step—we assume that the full $38\,mm$ width can be probed independently of $\phi$. Further below we will account for angle-dependent shadowing at the depth $z_0$ due to the limited probe aperture. Independent of a respective pivot, we use all nodes for echo simulation.

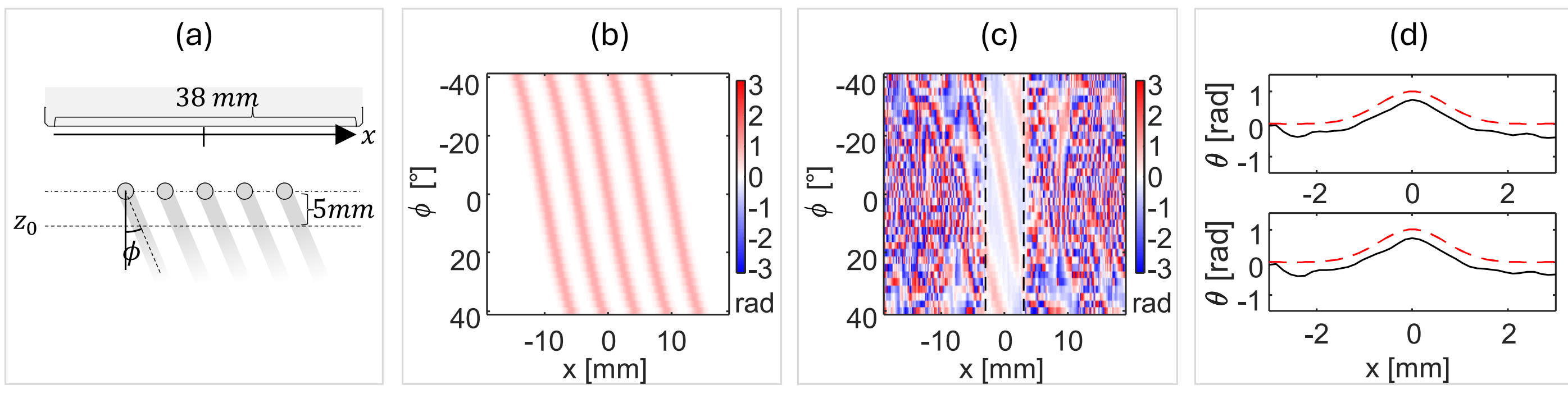

**Fig. 4:** (a) Simulation geometry with cylindrical contrast areas. (b) Ground truth phase delay $\theta$ as function of $x$ and $\phi$ at $z_0$. (c) Retrieved phase delay $\theta$ as function of $x$ and $\phi$ for a single pivot with $x_0 = 0$, for a beam radius $w_0 = 3\ mm$ (indicated by dashed vertical lines) but using the full $x$-interval for retrieval. (d) Profiles of retrieved (black solid line) vs. ground truth (red dashed line) of $\theta$ for $\phi = 0$, when using the full $x$-interval for retrieval (top) and when restricting retrieval to an interval $x_0 \pm w_0$ (bottom).

For the phase delay retrieval, we discretize $x$ with $\delta x = 0.15\ mm$ ($0.5\ \lambda_c$), resulting in 255 nodes. The grid resolution was chosen to fulfil the Nyquist condition for $\lambda_c$. From these nodes, the $x_n$ sampling the supports for different pivots are selected. We start with a single pivot at $x_0 = 0$, Gaussian $W_n$ with $1/e^2$ radius $w_0 = 3\ mm$, $L = 5$ layers, and $\phi_m$ ranging within $\pm 40°$ with a $2.5°$ spacing. For the retrieval we use 5 iterations, with the regularisation parameters $\gamma_\varepsilon^2$ and $\gamma_\theta^2$ set to 0.2 (empirical choice). To account for the lateral resolution limited by the element pitch, the phase delay is convolved in $x$ with a 3-tap Hann window (coefficients $[0.25, 0.5, 0.25]$). **Fig. 4c** shows the retrieved $\theta_\phi(x)$ when using the full $x$-interval (255 nodes) as support. Inside the interval $x_0 \pm w_0$ indicated by dashed lines, the retrieved phase delay agrees qualitatively with the ground truth. For a quantitative comparison, **Fig. 4d**, top, shows the retrieved phase delay profile for $\phi = 0$ vs. the ground truth inside the interval $x_0 \pm w_0$. Apart from an offset, the agreement between these profiles demonstrates the feasibility of phase delay retrieval. The offset will be discussed further below. In a next step, we restrict the sampled support to $N = 61$ nodes $x_n$ inside the interval $x_0 \pm 1.5 w_0$. **Fig. 4d**, bottom, shows the resulting phase delay profile, and **Fig. 5a** shows the 2D map. These results demonstrate that the restriction of the support has negligible influence on the retrieved $\theta$ inside the interval $x_0 \pm 1.5 w_0$. When restricting $x_n$ further (not shown), the significant influence of unaccounted reflectivity and phase delay outside the interval edges leads to increased errors.

Next, we illustrate the relationship between beamwidth and the angular spacing required for phase delay retrieval. **Fig. 5b** shows the result when quadrupling $w_0$ compared to **Fig. 5a** ($12\ mm$ instead of $3\ mm$) but without adapting the angular spacing ($2.5°$). The structure of the phase delay is hardly visible due to strong noise. However, when compensating the 4-fold larger $w_0$ with a 4-fold smaller angular spacing ($0.625°$ instead of $2.5°$, **Fig. 5b**), the same quality is obtained as in **Fig. 5a**., however, inside a larger interval $x_0 \pm w_0$. The problem of this approach to enlarging the FoV is that—due to the factor 4 increase in number of $x_n$ and $\phi_m$ , the size of the matrices $\mathbf{\Theta}$ and $\mathbf{E}_{m2}$ in **Eq. 9** and **11** scales by a factor $4^2 \times 4$, the computational cost for the products $\mathbf{\Theta}^*\mathbf{\Theta}$ and $\mathbf{E}_{m2}^*\mathbf{E}_{m2}$ scales by a factor $4^2 \times 4^2$, and the computational cost for the matrix inversion scales by a factor $4^3$ (assuming Gaussian elimination). As we will see further below, it also leads to artifacts when using realistic data due to the violation of the small-angle approximation.

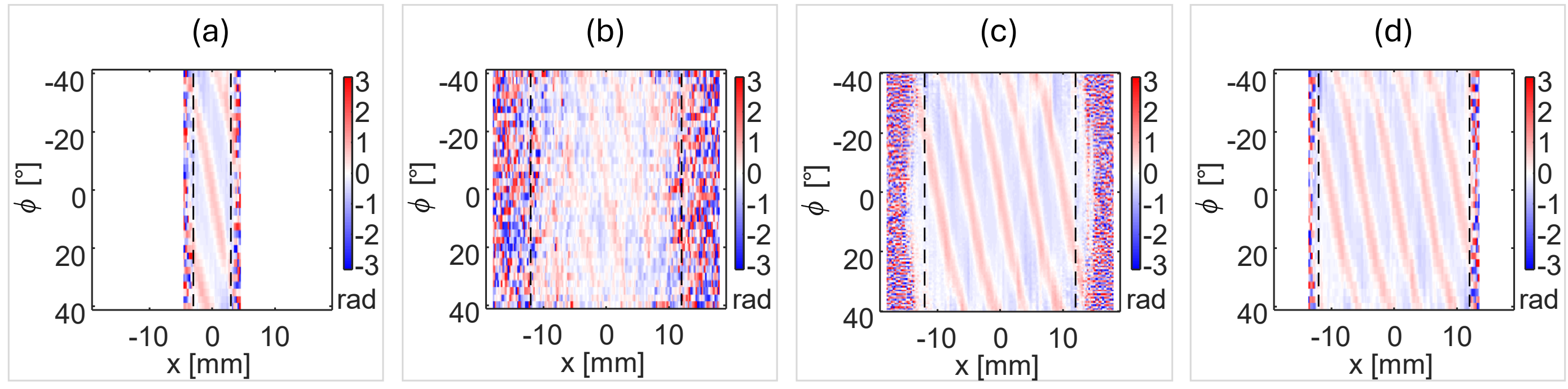

**Fig. 5:** (a) Retrieved $\theta$ when restricting the retrieval to the interval $x_0 \pm w_0$, for a single pivot with $x_0 = 0$, $w_0 = 3\ mm$, and angular spacing $\delta\phi = 2.5°$; (b) same as (a) but for a larger $w_0 = 12\ mm$; (c) same as (b) but with reduced $\delta\phi = 0.625°$. (d) Combination of data from 7 pivots spaced by $w_0$ for $w_0 = 3\ mm$ ($x_0 = [-9, \dots, 9] mm$), covering the same area as in (c) but using $\delta\phi = 2.5°$. The vertical dashed lines indicate the lateral limits determined by $x_0$ and $w_0$.

To avoid these problems, we keep $w_0 = 3\,mm$ and $\delta\phi = 2.5°$ but combine data for 7 pivots with $x_0$ spaced by $3\,mm$. A spatially weighted sum (using the profiles $W^2$ as weights) of the 7 phase delay profiles (shown in **Fig. 5d**) covers the same FoV as in **Fig. 5c,** but the computational cost scales only by the number of pivots (factor 7) compared to the result in **Fig. 5a**.

We mentioned in the Theory section that simultaneous phase delay retrieval from multiple independent reflectivity layers reduces noise, and we have chosen the number of layers $L = 5$ because this value is reasonable in a realistic scenario (see further below). **Fig. 6a** shows the result from **Fig. 5d** (combination of 7 pivots), but using $L = 2$. Compared to **Fig. 5d** this result shows elevated noise. Note that this is the lowest $L$ for which the structure of the phase delay is still recognizable. To quantitatively investigate how the retrieval quality depends on $L$, we define the retrieval error as the root-mean-square error (RMSE) between the retrieved $\theta_{\phi m}(x_n)$ and the ground truth $\theta_{\phi m}^{true}(x_n)$:

$$RMSE = \sqrt{\langle \mathcal{R}\left(\theta_{\phi m}(x_n) - \theta_{\phi m}^{true}(x_n)\right)^2 \rangle_{m,n}}, \quad (14)$$

where the brackets <> refer to the mean over $\phi_m$ and $x_n$, for $x_n$ inside $\pm 9\,mm$. The function $\mathcal{R}$ indicates a transformation of the data before computing the mean square value. This transformation is described further below in **Eq. 15**, and—for each $x_n$—it projects $\theta_{\phi m}(x_n)$ and $\theta_{\phi m}^{true}(x_n)$ to the space of retrievable profiles. The RMSE is normalized (nRMSE) by the RMS value of the ground truth after the same transformation, i.e., of $\mathcal{R}\left(\theta_{\phi m}^{true}(x_n)\right)$.

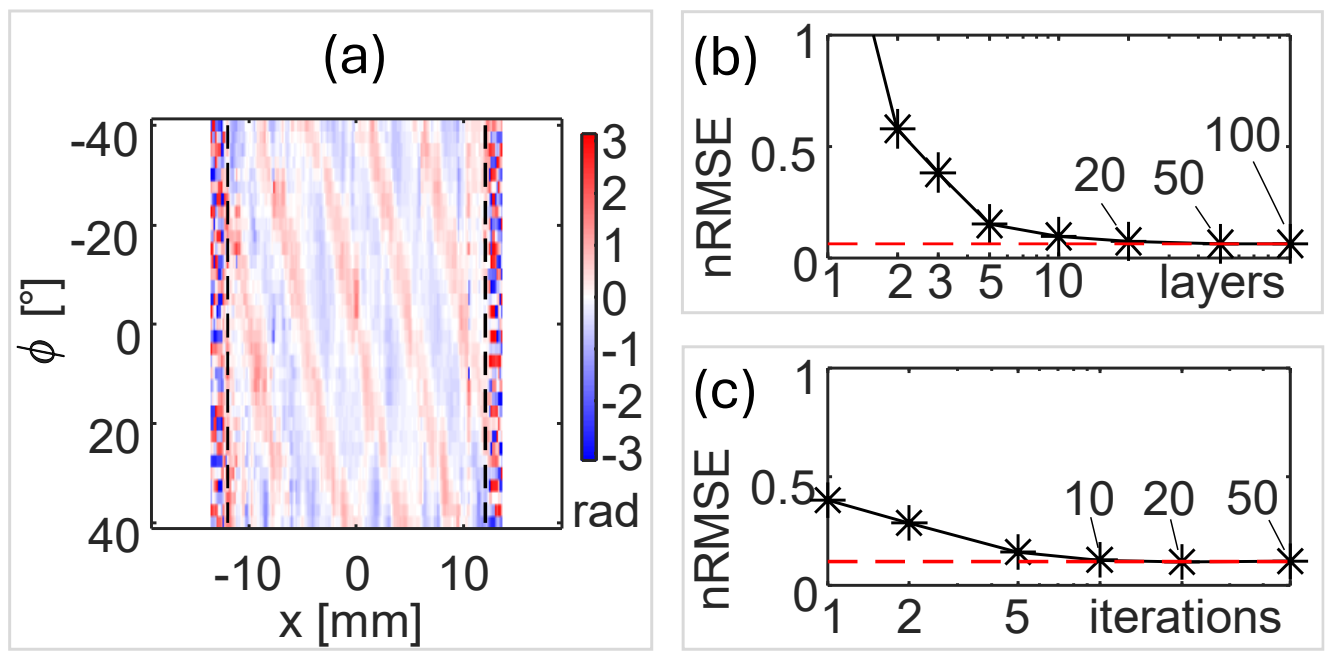

**Fig. 6:** (a) Phase delay retrieved from $L = 2$ reflectivity layers. (b) Convergence of retrieval error (nRMSE) for increasing number of reflectivity layers, for 5 iterations. (c) Convergence of nRMSE for increasing number of iterations, for 5 reflectivity layers. In (b) and (c), the dashed red line indicates the minimum achieved nRMSE value.

**Fig. 6b** shows the nRMSE as a function of number $L$ of reflectivity layers. The graph confirms that a minimum of 2 layers is required for sufficient contrast (nRMSE below 1). It also shows that—even though 5 layers lead to good contrast—substantial further reduction of nRMSE would be possible for $L > 20$. **Fig. 7a** ($L = 50$) illustrates the improved contrast compared to **Fig. 5d**. Unfortunately, this is not achievable in practice where the number of independent reflectivity layers is limited by the desired axial resolution of the phase delay (e.g., $1\,mm$), divided by the axial resolution $\delta z$ of the reflectivity (e.g., $0.2\,mm$). Our results were so far based on 5 iterations of **Eq. 9** and **Eq. 11**. In **Fig. 6c** we support this choice by the nRMSE for various iteration counts: with 5 iterations the nRMSE has nearly converged and the negligible improvement with further iterations does not justify the additional computational cost.

In **Fig. 4** we pointed at a negative bias between the retrieved and ground truth phase delay. This bias is confirmed in **Fig. 7a**, where 13 pivots spaced by $\Delta x_0 = 3\,mm = w_0$ are combined to cover the full probe aperture length. Here, for ease of comparison, 50 reflectivity layers are used to minimise noise. It is straight-forward to assume that the bias is related to the well-known ambiguity between mean phase delay (representing SoS) and reflectivity phase (representing axial reflector position): for this reason, only relative phase delay values can be retrieved. **Fig. 7d**, top, exemplarily shows the retrieved phase delay as a function of $x$ for a single angle $\phi = -37.5°$ (red line). The dashed black line shows the ground truth after subtracting—for each $x$—the mean along $\phi$. This "mean subtraction model" explains the observed bias quite well except for small deviations (grey arrows). However, when reducing $z_0$ from $5\,mm$ to $2\,mm$ (thus reducing the lateral shift of the phase delay profiles with increasing $\phi$), the retrieved phase delay (2D map in **Fig. 7b**) qualitatively changes compared to **Fig. 7a**, and the profile for $\phi = -37.5°$ (**Fig. 7d**, bottom) does not match the mean subtraction model. The reason is found in a second ambiguity: the one between (i) lateral reflector positions and (ii) the effective beam-steering angle due to a linear trend in the dependence of phase delay on $\phi$ (corresponding to a transversal gradient of SoS). To account for both ambiguities, we write $\theta_{\phi m}(x_n)$ as a sum of a constant $\theta^{(0)}(x_n)$ (cannot be retrieved due to the first ambiguity), a trend $\theta^{(1)}(x_n) \cdot \phi$ (cannot be retrieved due to the second ambiguity), and a retrievable part $\hat{\theta}_{\phi m}(x_n)$. Our refined model of the relationship between $\hat{\theta}_{\phi m}(x_n)$ and $\theta_{\phi m}(x_n)$ is thus expressed as a linear projection operator $\mathcal{R}$:

$$\hat{\theta}_{\phi m}(x_n) = \mathcal{R}\left(\theta_{\phi m}(x_n)\right)$$
$$= \theta_{\phi m}(x_n) - \theta^{(0)}(x_n) - \theta^{(1)}(x_n) \cdot \phi_m \;, \quad (15)$$

where $\theta^{(0)}(x_n) = \langle \theta_{\phi m}(x_n) \rangle_m$ ,

and $\theta^{(1)}(x_n) = \frac{\langle \theta_{\phi m}(x_n) \cdot \phi_m \rangle_m}{\langle {\phi_m}^2 \rangle_m}$ .

**Fig. 7c** shows $\hat{\theta}_{\phi m}(x_n)$ for the scenario in **(b)**: it nicely reproduces the retrieved phase delay. The graphs in **Fig. 7d** (solid black lines) confirm the accuracy of the refined model: it not only improves the prediction for result **(a)** (top) where the mean subtraction model was inaccurate (gray arrows), but it accurately matches the profile for **(b)** (bottom) where the mean subtraction model failed.

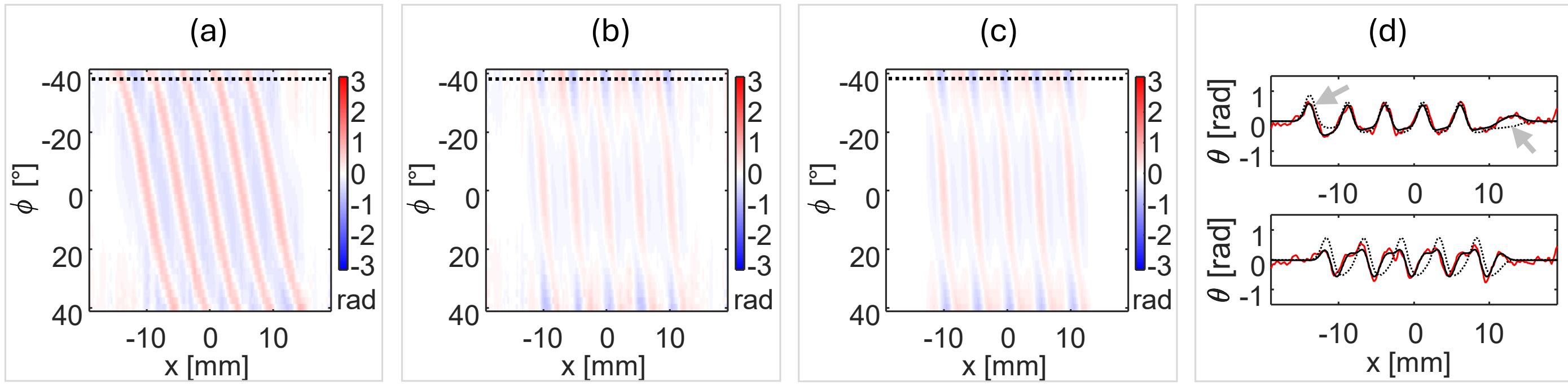

**Fig. 7:** Comparison of retrieved and predicted phase delay. All results are based on 50 reflectivity layers and 5 iterations. (a) Retrieved phase delay for the ground truth in **Fig. 4b** for $z_0$ 5 $mm$ below cylinders. (b) Retrieved phase delay for $z_0$ 2 $mm$ below cylinders. (c) Prediction of result (b) based on ground truth, using the refined model. (d) Profiles for $\phi = -37.5°$ (dashed lines in (a) to (c)), for $z_0$ 5 $mm$ (top) and 2 $mm$ below cylinders (bottom). Red line: ground truth; dotted black line: mean subtraction model; solid black line: refined model.

3.2 a-CUTE phase delay retrieval in realistic simulations

In the previous sub-section, we demonstrated the theoretical feasibility of retrieving highly resolved transversal profiles of phase delay from the complex echo amplitudes. However, these results were obtained with idealized data, simulated by the forward model itself which included various assumptions that can be problematic in practice. Next, we investigate how well the technique transfers to more realistic data that combines a wave propagation simulation with delay-and-sum beamforming.

For comparability with the previous section, our simulation follows the scenario set out in **Fig. 4a**: 5 cylinders with 1 $mm$ radius and a 40 $m/s$ SoS difference (corresponding to 1 $rad$ phase delay at 5 $MHz$) are located inside a uniform background SoS (1540 $m/s$), with a transversal spacing of 5 $mm$ at a depth of 5 $mm$. The same probe parameters are assumed (5 $MHz$ centre frequency, 0.3 $mm$ pitch, 128 elements). A full-matrix capture (FMC) is simulated using a combination of the hybrid angular spectrum method (to model the influence of SoS on wave propagation) with the 1st order Born approximation (to model echo generation), resulting in raw radio frequency (RF) signals $S(p, q, t)$ where $p$, $q$ index the Tx and Rx elements. Note that—in the definition of the PSF model (**Eq. 5**)—the transversal component of the $k$-vector of the Tx/Rx beam, $k_{x,\phi} = k \sin\phi$, implicitly assumes that the complex amplitude is right-rotating with propagation distance, and thus left-rotating with time. To match this convention, the *complex conjugate* of the Hilbert transform is used to translate the RF signals to left-rotating complex RF (CRF) signals $\tilde{S}(p, q, t)$. The CRF FMC signals are beamformed to collimated Gaussian Tx/Rx beams using delay-and-sum with linear Tx/Rx steering delays and Gaussian apodization functions $W(x)$ (1/e radius $w_0 = 3\ mm$), for Tx/Rx angles $\phi_m$ ranging within $\pm 40°$ with a 2.5° spacing. The beamforming SoS is chosen identical to the background, $c_b = 1540$ m/s. Note that a positive SoS contrast leads to a shift of the wavefield into the direction of propagation, corresponding to a negative phase delay (earlier arrival). For ease of comparison, we will thus display the phase delay with flipped sign to match the positive phase delay used in the previous results.

For simplicity, we have so far assumed that the FoV can be probed on the full width of the array aperture. In practice, however, the aperture length limits the area to/from which US power can be transmitted/received depending on $\phi$. We therefore modify the forward model **Eq. 7** by accounting for the acoustic shadows with a mask $\chi(x, \phi)$:

$$u_{m,m'} = \sum_{n=1}^{n=N} \chi_{m,n} \cdot \chi_{m',n} \cdot W_n{}^2 \cdot e^{i(k_m + k_{m'})x_n} \cdot \Theta_{m,n} \cdot \Theta_{m',n} \cdot \tilde{\varepsilon}_n \,, \tag{16}$$

where $\chi_{m,n} = \chi(x_n, \phi_m)$ is 0 inside the acoustic shadow and 1 inside the detected area. Accordingly, the prediction model for the phase delay is adapted to account for $\chi$:

$$\hat{\theta}_{\phi_m}(x_n) = \theta_{\phi_m}(x_n) - \theta^{(0)}(x_n) - \theta^{(1)}(x_n) \cdot \hat{\phi}_m \,, \tag{17}$$

where $\theta^{(0)}(x_n) = \frac{\langle \chi_{m,n} \cdot \theta_{\phi_m}(x_n) \rangle_m}{\langle \chi_{m,n} \rangle_m}$ ,

$\hat{\phi}_m = \phi_m - \frac{\langle \chi_{m',n} \cdot \theta_{\phi_{m'}}(x_n) \rangle_{m'}}{\langle \chi_{m',n} \rangle_{m'}}$ ,

and $\theta^{(1)}(x_n) = \frac{\langle \chi_{m,n} \cdot \theta_{\phi_m}(x_n) \cdot \hat{\phi}_m \rangle_m}{\langle \chi_{m,n}{}^2 \cdot \hat{\phi}_m{}^2 \rangle_m}$ .

**Fig. 8** shows the retrieved phase delay for a single $z_0 = 10\ mm$ (i.e., 5 $mm$ below the cylinder depth) and 5 reflectivity layers spaced by $\delta z = 0.2\ mm$. **Fig. 8a** is the result when using a single pivot with $x_0 = 0$ and a large beamwidth $w_0 = 12\ mm$ together with an angular spacing $\delta\phi = 0.625°$. Compared to the corresponding result in **Fig. 5c** it shows elevated noise (white arrowheads) and lack of detail (black arrowhead) in areas away from $x_0$.

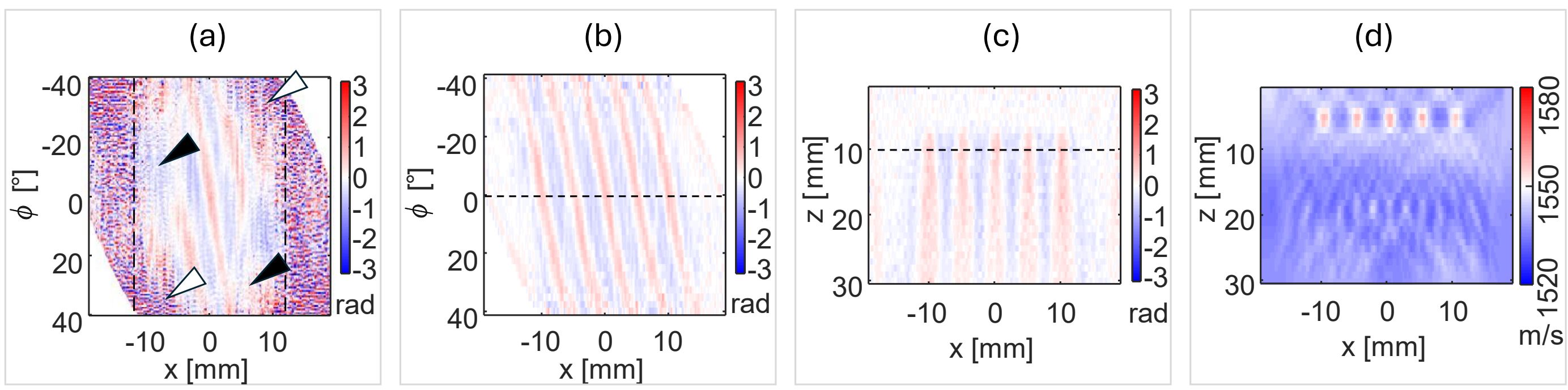

**Fig. 8:** (a) and (b) Phase delay retrieved from realistic simulations of the scenario in Fig. 4a for single $z_0 = 10\ mm$ ($5\ mm$ below cylinders), for (a) single $x_0 = 0$ with $w_0 = 12\ mm$ and angular spacing $\delta\phi = 0.625°$, and (b) combining 13 pivots with $x_0$ spaced by $3\ mm$ with $w_0 = 3\ mm$ and angular spacing $\delta\phi = 2.5°$. 5 reflectivity layers were used with $\delta z = 0.2\ mm$. (c) Phase delay using same parameters as in (b) but for multiple $z_0$ covering $30\ mm$ depth range, exemplified for $\phi = 0°$. (d) SoS map retrieved from the phase delay data.

A notable difference of **Fig. 8b** compared to **Fig. 5d** is the increased noise outside $\phi = \pm 25°$, which may be attributed to the small-angle approximation but also to grating lobes (element pitch equals the centre wavelength).

**Fig. 8b** is the result when using a small beamwidth $w_0 = 3\ mm$ with angular spacing $\delta\phi = 2.5°$, but combining 11 pivots with $x_0$ spaced by $3\ mm$ to cover the full aperture length. The structure of the phase delay profiles is strongly improved, and the quality of this result is close to what was obtained in the corresponding idealized scenario in **Fig. 5d**. The apparent reason for the reduced quality in **Fig. 8a** compared to **Fig. 5c** and **Fig. 8b** is the violation of the small-angle approximation due to the large angle limits of $(\varphi, \psi)$: the tails of the PSF can rotate relative to the $x$-axis by more than the axial width $\delta z$ of the profile $T^2(.)$, leading to a dependence of $\tilde{\varepsilon}(x')$ on $(\varphi, \psi)$. This problem is strongly reduced for the smaller PSF radius in **Fig. 8b**.

For imaging, the procedure is repeated for $z_0$ ranging from 0 to $30\ mm$ in $1\ mm$ steps. **Fig. 8c** shows the retrieved phase delay as a function of $x$ and $z$ exemplarily for angle $\phi = 0°$. This result clearly shows the axial streaks of phase delay spreading from the depth of the cylinders downwards.

## 3.3 a-CUTE SoS retrieval in realistic simulations

From such phase delay maps, the spatial distribution of the SoS can be retrieved. For illustration, here we implement a preliminary SoS retrieval that closely follows the methodology used in dToF-CUTE [20]: we pseudo-invert a linear matrix operator $\mathbf{M}$ that relates the slowness deviation, defined as $\sigma(x,z) = 1/c(x,z) - 1/c_b$, to the measured phase delay. This operator can be seen as a product of two matrices. A first matrix, $\mathbf{A}$, relates the slowness deviation to the true phase delay. The second matrix, $\mathbf{R}$, describes the relation between the true and the measured phase delay values, with matrix elements determined by the prediction model **Eq. 17**. The pseudo-inverse is obtained as:

$$\mathbf{M}^{\dagger} \stackrel{\text{def}}{=} \left(\mathbf{M}^T\mathbf{M} + \gamma_x^2 \boldsymbol{D}_x^T \boldsymbol{D}_x + \gamma_z^2 \boldsymbol{D}_z^T \boldsymbol{D}_z\right)^{-1} \mathbf{M}^T, \tag{18}$$

where $\mathbf{M} = \mathbf{RA}$. The regularisation matrices $\boldsymbol{D}_x$ and $\boldsymbol{D}_z$ implement two-tap 1st order finite differences (coefficients $\pm 1$) along $x$ and $z$ to enforce smooth solutions, and $\gamma_x^2$ and $\gamma_z^2$ are the respective regularisation parameters. The slowness deviation values are estimated from the retrieved phase delay values as:

$$\boldsymbol{\sigma} = \mathbf{M}^{\dagger}\boldsymbol{\theta}, \tag{19}$$

and then converted to SoS $c(x,z)$. To limit computational cost by reducing matrix sizes, we coarsen the grid resolution of the retrieved 3D phase delay data for SoS inversion: the $x$-axis by a factor of two (to $0.3\ mm$ step size), and the angles by a factor 4 ($10°$ step size). As mentioned above, phase delay profiles are noisy outside angles $\phi = \pm 25°$. Analogue to previous publications, we therefore restrict the SoS inversion to the data within this angle range. **Fig. 8d** shows the SoS map when using the above-described SoS inversion method, for $\gamma_x$ and $\gamma_z$ set to 2 and 1, respectively. The 5 cylinders are well resolved, demonstrating the feasibility of SoS retrieval following amplitude-based phase delay retrieval.

Finally, we illustrate the benefit of a-CUTE compared to the classical dToF-CUTE. For dToF-CUTE, echo amplitudes $u(\ldots)$ are beamformed for pivots spaced by $\delta x_0 = 0.9\ mm$ but otherwise using the same parameters $w_0 = 3\ mm$, $\delta z = 0.2\ mm$, $\delta z_0 = 1\ mm$, $\delta\phi = 2.5°$. The rest of the methodology closely follows previous publications [17,20]: (i) For each $(x_0, z_0 + l\delta z)$, The Hermitian product is computed between image pairs $u(\ldots, \varphi_{m1}, \psi_{m2})$ and $u(\ldots, \varphi_{m1+1}, \psi_{m2-1})$ that have the same mid-angle $\frac{1}{2}(\varphi_{m1} + \psi_{m2}) = \frac{1}{2}(\varphi_{m1+1} + \psi_{m2-1})$, and averaged over $l$ to get a zero-lag cross-correlation value for $(x_0, z_0)$. (ii) The phase shift is computed as the argument of this zero-lag cross-correlation and summed over successive image pairs to obtain the compound phase shift between angle pairs $(\varphi, \psi)$ with $\varphi/\psi \in \{-25°{:}\,10°{:}\,25°\}$. This results in 25 phase shift maps (one per successive angle pairs with the same mid-angle), sampled with a spatial resolution of $0.9\ mm$ ($x$) by $1\ mm$ ($z$). The SoS retrieval from these maps is performed like in **Eq. 18** and **Eq. 19**, with the difference that $\boldsymbol{\theta}$ now corresponds to phase shift, and $\mathbf{R}$ governs the transformation from the true phase delay to phase shift, see reference [20] for details.

For the comparison of a-CUTE and dToF-CUTE, we use a digital phantom which contains a row of 6 cylinders at $5\ mm$ depth arranged with increasing spacing: $2\ mm$, $3\ mm$, $4\ mm$, $5\ mm$, and $8\ mm$. Each cylinder has a $1\ mm$ diameter and a $180\ m/s$ SoS difference relative to the $1540\ m/s$ background. **Fig. 9a** and **9b** show examples of the phase delay (a-CUTE) and phase shift (dToF-CUTE). These maps already illustrate the quality difference between the two approaches: while a-CUTE retrieves phase delay maps where the delay from individual cylinders can be distinguished, this is hardly possible in dToF-CUTE due to the substantial transversal blurring. In addition, the dToF map shows salt and pepper-like noise that is caused by unresolved interference of the wavefields from the different cylinders. **Fig. 9c**, top, shows the a-CUTE SoS map obtained with the same parameters as **Fig. 8d**. The contrast of the cylinders is well visible, however, the closest two are not resolved even though their distance ($2\ mm$) is larger than the cylinder diameter ($1\ mm$). Interestingly, this situation changes when restricting the SoS inversion to a $10\ mm$ depth range (**Fig. 9c**, bottom): all cylinders are resolved, and this with improved contrast. **Fig. 9f**, subpanel (f-c), shows a lateral SoS profile along the dashed line. The smaller SoS difference relative to the background (around $70\ m/s$), compared to the ground truth ($180\ m/s$), can be explained with the partial volume effect as the retrieved diameter of the contrast areas is larger than the actual $1\ mm$: around $3\ mm$ in $z$ and $1\ mm$ in $x$. In the SoS map retrieved from a $30\ mm$ depth range, the lower part contains artefactual noise. To a lesser extent, this noise could already be seen in **Fig. 8d**. The blue profile in **Fig. 9f**, subpanel (f-c) shows—for each $x$—the mean SoS $\pm$ its standard deviation along $z$ inside the area denoted by the dashed rectangle in **Fig. 9c**. We hypothesize that this noise stems from the disagreement between the wave nature of ultrasound and the straight-ray approximation used for SoS retrieval, which is more pronounced for shorter-scale spatial fluctuations of SoS. We further hypothesize that the reduced resolution and contrast of the cylinders for the $30\ mm$ compared to the $10\ mm$ retrieval depth is caused by the same disagreement, as misinterpretation of data at depth (where interference becomes increasingly important) would be compensated by erroneous SoS retrieval in the overlying areas.

**Fig. 9d** shows the SoS map retrieved with dToF-CUTE with regularisation parameters ($\gamma_x = 4, \gamma_z = 2$). This map shows the cylinders with notably reduced spatial resolution compared to a-CUTE. This is especially seen in the inability to distinguish the closest three cylinders (see also profile in **Fig. 9f**, subpanel (f-d)). Here, limiting the SoS retrieval to a smaller depth range does not improve resolution. Below the cylinders, strong artefactual noise is observed. Note that this noise is strongest below the non-resolved cylinders (see also **Fig. 9f**, subpanel (f-d)), supporting the hypothesis that this noise is a secondary result to the lack of resolution. Stronger regularisation reduces the level of artifacts, exemplarily shown in **Fig. 9e** for ($\gamma_x = 12, \gamma_z = 6$) (profiles in **Fig. 9f**, subpanel (f-e)), however, this also reduces the cylinder contrast, demonstrating that a-CUTE and dToF-CUTE have different contrast-resolution trade-offs.

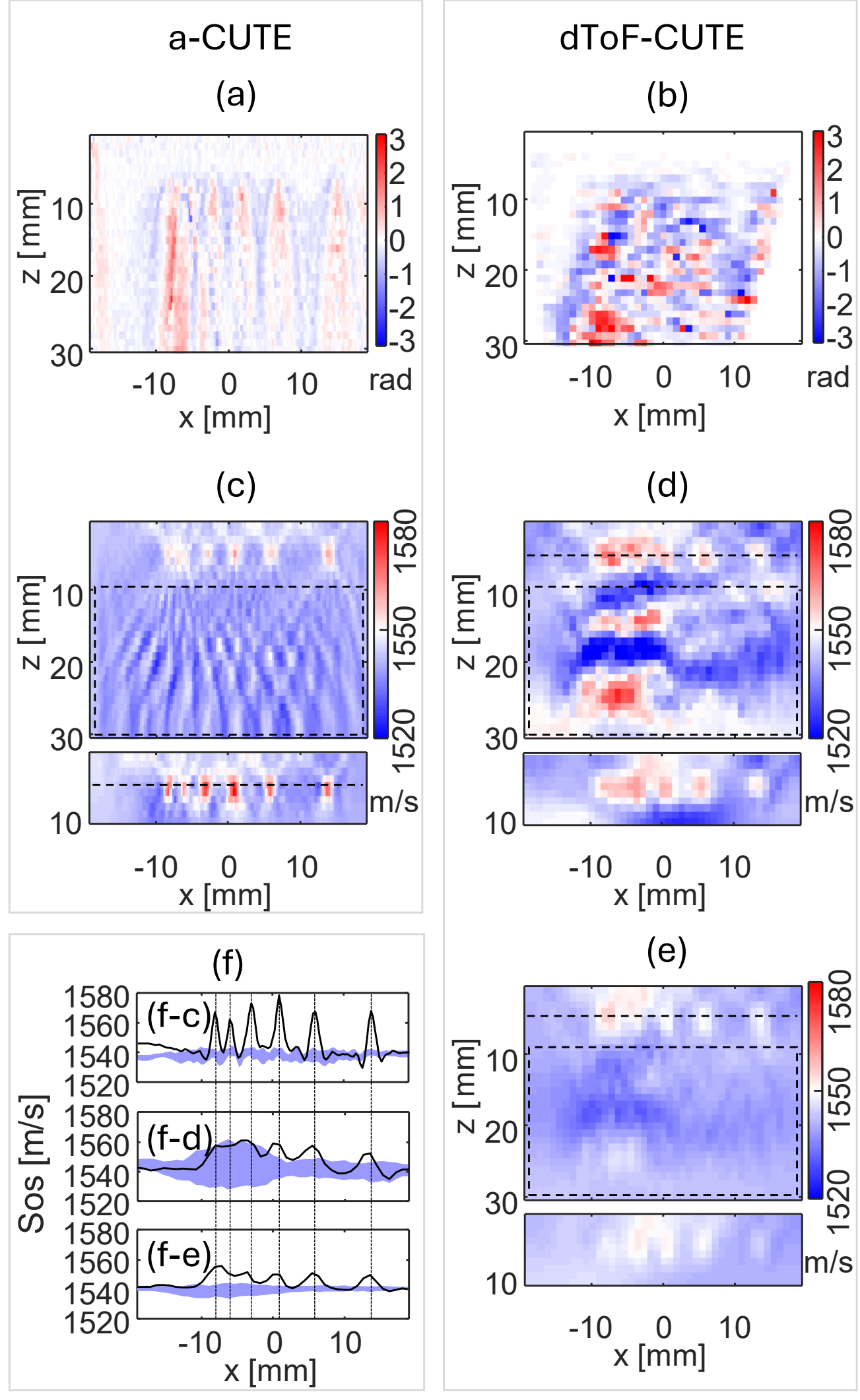

**Fig. 9:** Comparison of a-CUTE and dToF-CUTE, for 6 cylinders ($1\ mm$ diameter, $180\ m/s$ SoS difference) arranged with increasing spacing: $2\ mm$, $3\ mm$, $4\ mm$, $5\ mm$, and $8\ mm$: (a) a-CUTE phase delay, exemplified for $\phi = 0°$. (b) dToF-CUTE phase shift map, exemplified for the transition from the Tx/Rx angle $(-5, -5)$ to $(5, -15)$. (c) a-CUTE SoS map ($\gamma_x = 2, \gamma_z = 1$). (d) and (e) dToF-CUTE SoS maps using different regularisation weighs, d: ($\gamma_x = 4,\ \gamma_z = 2$) and e: ($\gamma_x = 12,\ \gamma_z = 6$). SoS maps were retrieved for $30\ mm$ (top) and $10\ mm$ depth range (bottom). (f-c) to (f-e) Transversal profiles (black lines) of SoS taken along the dashed lines in (c) to (e), together with transversal profiles of the mean SoS $\pm$ its standard deviation (blue) along $z$ inside the dashed rectangles in (c) to (e). The dashed vertical lines indicate the true cylinder positions.

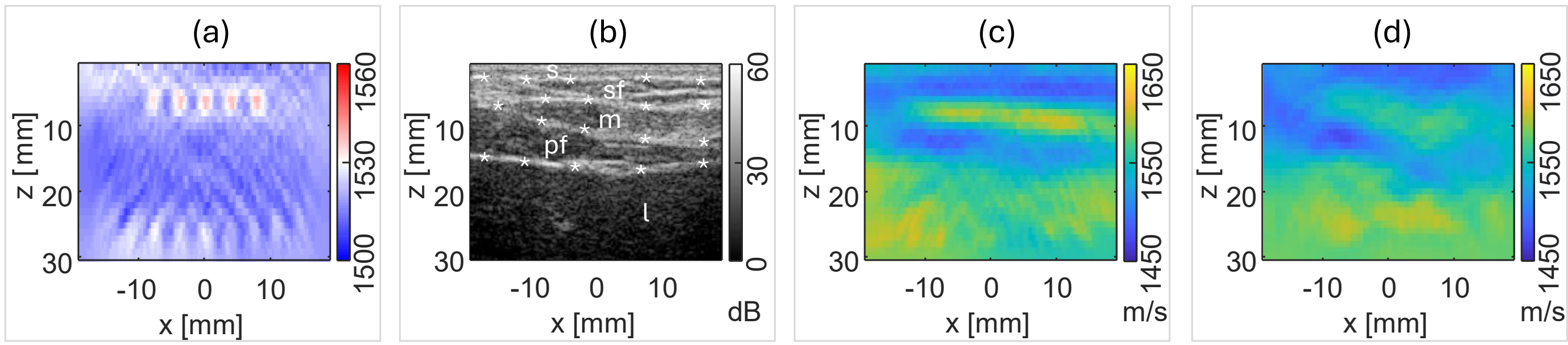

**Fig. 10:** Physical results. (a) a-CUTE SoS map of gelatine phantom containing a row of 5 cylinders with 2 $mm$ diameter and 45 $m/s$ SoS difference relative to a uniform background. (b) B-mode image and (c) a-CUTE SoS map of the abdomen of a healthy volunteer. The interfaces between different tissues are marked with asterisks in (b), and the tissues are indicated (s: skin; sf: subcutaneous fat; m: muscle; pf: peritoneal fat; l: liver). (d) dToF-CUTE derived from the same data.

3.4 Proof of principle of a-CUTE in physical data

To demonstrate a-CUTE in physical data, we design a phantom containing a row of 5 cylinders with $2\,mm$ diameter placed at $5\,mm$ depth inside a uniform background. The background and cylinders are made of $10\,wt\%$ and $20\,wt\%$ gelatine, respectively, in water (Geistlich Spezial Gelatine, health and life AG, Switzerland), with $2\,wt\%$ baking flower added to provide diffuse echogenicity. Reference SoS values were obtained on separate samples of the same mixtures using through-transmission ToF measurements, yielding $(1521.7 \pm 3.5\,)\,m/s$ and $(1567.6 \pm 3.5)\,m/s$, respectively, corresponding to a SoS difference of $(45.9 \pm 4.9)\,m/s$ comparable to the $40\,m/s$ that were used for the simulation in **Fig. 8**. As a first investigation whether a-CUTE can already give reasonable results in real tissue, we additionally show an example of abdominal a-CUTE (healthy adult, informed consent, in accordance with The Code of Ethics of the World Medical Association (Declaration of Helsinki) for experiments involving humans). We acquired FMC data using an L7-4 linear array probe (ATL Philips, WA, USA) featuring 128 elements at $0.3\,mm$ pitch with $5\,MHz$ centre frequency, connected to a Vantage128 research ultrasound system (Verasonics, Inc., WA, USA). The Tx voltage was 31V, and the sampling frequency $19.52\,MHz$.

**Fig. 10a** is the retrieved SoS map of the phantom. This result shows an excellent qualitative agreement of the cylinder's retrieved contrast distribution with the simulation result in **Fig. 8d**. The reason for the slightly increased background variations at the lower image edge compared to the simulation is not yet clear, however, they have little influence on the cylinder contrast.

**Fig. 10b** to **10d** show the *in vivo* result. The spatial distribution of SoS on the a-CUTE SoS map (**10c**) agrees well with the tissue structures seen in B-mode (**10b**), with high SoS (around 1600 m/s) inside the muscle and liver, and low SoS (around 1500 m/s) inside the fat layers. Even though these values may not be quantitatively correct (see Conclusion), they are in qualitative agreement with expectations [39]. For comparison, **Fig. 10d** is the dToF-CUTE SoS map. Whereas the amplitude of artefactual SoS variations inside the liver looks similar (by the choice of the regularisation strength), the interfaces between the peritoneal fat layer and the liver appear substantially distorted compared to the ground truth and the a-CUTE result. Note that this comparison was obtained with adapted parameters compared to simulation/phantom study: (i) to reduce noise, the pivot spacing for a-CUTE was $\delta x_0 = 0.9\,mm$ (same as for dToF-CUTE), and the angular step for retrieval was $5°$ instead of $10°$, resulting in an averaging effect for both techniques; (ii) to provide the visually most pleasing result for both techniques, the regularisation parameters were adjusted to $(\gamma_x = 5, \gamma_z = 0.5)$ for a-CUTE and $(\gamma_x = 6, \gamma_z = 3)$ for dToF-CUTE. We would like to point out that—given the absence of a ground truth—a parameter optimization is beyond the scope of this study. The exemplary result shall merely confirm that a-CUTE can work *in vivo*.

## 4. Conclusion

In this study we have demonstrated the feasibility of a-CUTE in simulations and physical phantom data, and that this approach carries huge potential for improving spatial resolution and contrast of SoS maps compared to classical dToF-CUTE in samples with heterogeneous SoS. The exemplary *in vivo* results confirm this finding.

The proposed algorithm is a first step towards a-CUTE, and various improvements will be necessary to fully leverage its increased reliability compared to dToF-CUTE. (i) We used the straight-ray approximation to derive SoS maps from the retrieved phase delay profiles. However, the detail level in these profiles suggests that diffraction/interference effects are resolved to an extent that allows to employ more accurate models of wave propagation for SoS retrieval for further improved resolution and reduced artifacts. The same as in through-transmission UT [40], this may be implemented computationally efficient by solving a finite frequency linear inverse problem. (ii) While dToF-CUTE employs relative measurements of phase shift between angle pairs, the proposed a-CUTE retrieves phase delays for individual angles and is thus more susceptible to phase aliasing (cycle skipping). Techniques are required to avoid this ambiguity, e.g., via an iterative approach where the retrieved SoS in superficial areas is used for disambiguation of phase delay in deeper areas. (iii) In areas with a large dynamic range of echo intensity, the rotating sidelobes of strong echoes when seen from different angles may be confounded with aberration-related phase delays. This problem was solved in dToF-CUTE by the common mid-angle approach [20], and similar strategies may be needed in a-CUTE. (iv) Even though it may be trivial to assume that phase delay directly represents ToF, a more elaborate model will be needed for quantitative accuracy as was the case for dToF-CUTE [20]. For this purpose—as an alternative to an analytic solution—a machine learning approach could be employed, to either learn the forward model similar to [41] or the pseudo-inverse for SoS retrieval similar to [42]. (v) For computational efficiency, the phase delay retrieval was based on a small-angle approximation. It is astonishing how little the results were affected by the actually very large angle range, however, the problems described above (large dynamic range artifacts, lack of quantitativeness) may be reduced with a more realistic model that takes PSF rotation into account.

Finally, any progress in the characterization and interpretation of short-scale wave aberrations will directly benefit the accuracy and spatial resolution of attenuation mapping [43], where it can help disambiguation of amplitude fluctuations caused by attenuation from the ones caused by self-interference.

## Acknowledgments

This research did not receive specific funding. We cordially thank Dr. Naiara Korta Martiartu for helpful discussions.

**Table I: Glossary**

| Variable(s) | Description | Typical value, unit |
|---|---|---|
| $x, z$ | transversal/axial coordinates | $mm$ |
| $t$ | time | $\mu s$ |
| $(x_0, z_0)$ | pivot coordinates | $mm$ |
| $x', z'$ | coordinates relative to pivot | $mm$ |
| $\varphi, \phi$ | transmit (Tx)/ receive (Rx) angles | ° |
| $s(x_0, z_0, \varphi, \psi, t)$ | signal | unitless |
| $p_{\varphi/\psi}^{(x_0,z_0)}(x, z, t)$ | beamformed wavefields | unitless |
| $\varepsilon(x, z)$ | reflectivity | unitless |
| $t_b(x_0, z_0, \varphi, \psi)$ | anticipated round-trip time | $\mu s$ |
| $c_b$ | *a priori* (uniform) SoS | $1540\ m/s$ |
| $c(x, z)$ | spatial distribution of SoS | $m/s$ |
| $u(x_0, z_0, \varphi, \psi)$ | complex radio frequency (CRF)-mode map of echo amplitudes | unitless |
| $h_{\varphi,\psi}^{(x_0,z_0)}(x, z)$ | point-spread function (PSF) | unitless |
| $P_{\varphi/\psi}^{(x_0,z_0)}(x, z, t)$ | undisturbed wavefields | unitless |
| $H_{\varphi,\psi}^{(x_0,z_0)}(x, z)$ | undisturbed PSF | unitless |
| $\theta_{\varphi/\psi}(x, z)$ | aberration profile (real part: phase delay) | radians |
| $W_{\varphi/\psi}(x)$ | transversal beam profiles | Gaussian, unitless |
| $w_0$ | Gaussian profile $1/e^2$ radius | $3\ mm$ |
| $T(z)$ | axial pulse envelope | unitless |
| $f_c$ | temporal centre frequency | $5\ MHz$ |
| $k$ | spatial (centre) frequency | $20.4\ mm^{-1}$ |
| $k_{x,\varphi/\psi}$ | transversal spatial frequency | $mm^{-1}$ |
| $\delta z$ | axial width of $T^2(z)$ = distance between reflectivity layers | $0.2\ mm$ |
| $\tilde{\varepsilon}(x)$ | effective reflectivity | unitless |
| $\{x_n, n \epsilon [1{:}N]\}$ | set of discretized $x'$, size $N$ | $mm$ |
| $\delta x$ | transversal grid spacing | $0.125\ mm$ |
| $\{\phi_m, m \epsilon [1{:}M]\}$ | set of discretized $\varphi, \psi$, size $M$ | $[-40°{:}\ 2.5°{:}\ 40°]$ |
| $\phi_{max}$ | maximum absolute Tx/Rx angle | $40°$ |
| $\delta\phi$ | angular spacing | $2.5°$ |
| $L, l$ | number and index of reflectivity layers | 5 |
| $u_{m1,m2}, \Theta_{m,n}, \tilde{\varepsilon}_n, W_n, k_m$<br>$u_{m1,m2,l}, \tilde{\varepsilon}_{n,l}$ | discretized (for discrete angles and $x$ coordinate) $u$, $e^{i\theta}$, $\tilde{\varepsilon}$, $W$, $k_x$ | |
| $\mathbf{u}$, $\boldsymbol{\theta}$, $\boldsymbol{\varepsilon}$, $\boldsymbol{\Theta}$, $\mathbf{E}$, $\boldsymbol{\Theta}^{\dagger}$, $\mathbf{E}^{\dagger}$ | vectorized $u_{m1,m2}$, $\Theta_{m,n}$, $\tilde{\varepsilon}_n$, model matrices and their pseudo-inverses | |
| $\gamma_{\varepsilon}^2$, $\gamma_{\theta}^2$ | regularisation parameters for reflectivity/phase delay retrieval | 0.2 |
| $\Delta x$, $\Delta z$ | transversal/axial spacing of pivots | $3\ mm, 1\ mm$ |
| $\lambda_c$ | wavelength at centre frequency | $0.3\ mm$ |
| $\mathcal{R}(\cdots)$ | projection of $\theta_{m,n}$ to the space of retrievable profiles | |
| $\hat{\theta}_{m,n}$ | retrievable part of $\theta_{m,n}$ | |
| $\chi(x,\phi)$, $\chi_{m,n}$ | acoustic shadow mask | unitless |
| $\sigma(x,z)$ | slowness deviation | $\mu s/mm$ |
| $\gamma_x^2$, $\gamma_z^2$ | regularisation parameters for SoS retrieval | $rad \cdot mm/\mu s$ |